\documentclass[aps,prd,twocolumn,nofootinbib,superscriptaddress]{revtex4}
\usepackage{graphicx}% Include figure files
\usepackage{amsmath}
\def\lesssim{\mathrel{\mathpalette\vereq<}}
\def\gtrsim{\mathrel{\mathpalette\vereq>}}

\newcommand{\gsim}{\lower.7ex\hbox{$\;\stackrel{\textstyle>}{\sim}\;$}}
\newcommand{\lsim}{\lower.7ex\hbox{$\;\stackrel{\textstyle<}{\sim}\;$}}
\def\OO{{\cal O}}

\newcommand{\GeV}{\,\mathrm{GeV}}

\newcommand{\keV}{\,\mathrm{keV}}

\newcommand{\GUT}{{\text{GUT}}}

\def\unit{\relax{\rm 1\kern-.26em I}}

\newcommand{\eff}{{\text{eff}}}
\newcommand{\Max}{{\text{max}}}

\begin{document}

% Page numbers bottom-center
\pagestyle{plain}

\title{
\begin{flushright}
\mbox{\normalsize SLAC-PUB-12480}
\end{flushright}
\vskip 20 pt

Unification and Dark Matter in a Minimal Scalar Extension of the Standard Model}

\author{Mariangela Lisanti and Jay G. Wacker}
\affiliation{
SLAC, Stanford University, Menlo Park, CA 94025\\
Physics Department, Stanford University,
Stanford, CA 94305 
}

%\date{\today}

\begin{abstract}
The six Higgs doublet model is a minimal extension of the Standard Model (SM) that addresses dark matter and gauge coupling unification.  Another Higgs doublet in the {\bf 5} representation of a discrete symmetry group, such as $S_6$, is added to the SM.    The lightest components of the 5-Higgs are neutral, stable and serve as dark matter so long as the discrete symmetry is not broken.    Direct and indirect detection signals, as well as collider signatures are discussed.    The five-fold multiplicity of the dark matter decreases its mass and typically helps make the dark matter more visible in upcoming experiments. \end{abstract}

 \maketitle

%%%%%%%%%%%%%%%%%%%%%%%%%%%%%%%%%%%%%%%%%%%%%%%%%%%
%%%%%%%%%%%%%%%%%%%%%%%%%

\section{Introduction}

The Standard Model (SM) of particle physics has enjoyed great success in explaining physics below the electroweak scale, but it is unlikely to remain the sole description of nature up to the Planck scale.  The SM does not contain a viable cold dark matter candidate, such as a weakly interacting massive particle (WIMP), or satisfactorily address the issue of gauge coupling unification.  Both of these issues hint at new physics beyond the electroweak scale.  In addition, the SM must be fine-tuned to regulate the large radiative corrections to the Higgs mass and the cosmological constant.  %For example, the Higgs vacuum expectation value is unstable due to quantum loop corrections that are $\OO(\Lambda_{pl}^2)$.  Similarly, there is now strong experimental evidence for a small, non-zero cosmological constant, which is far below the theoretical value $\OO(\Lambda_{pl}^4)$ \cite{Spergel:2003}.If one sets aside the CC problem as a result of some unexplored aspect of quantum gravity, then
It is possible to obtain a natural Higgs from supersymmetry \cite{Dimopoulos:1981zb}, large extra dimensions \cite{Nima:1999}, technicolor \cite{Susskind:1979}, Randall-Sundrum models \cite{Randall:1999}, or the little Higgs mechanism \cite{Nima:2001}.   However,  these models do not address the magnitude of the cosmological constant, which appears to be more fine-tuned than the Higgs mass, and leads us to question the central role of naturalness in motivating theories beyond the Standard Model. 

An alternative approach is to explore non-natural extensions of the SM in which fine-tuning is explained by environmental selection criteria \cite{Hogan:1999wh}.  Weinberg noted that if the cosmological constant was much larger than its observed value, galaxy formation could not occur \cite{Weinberg:1987}.  Additionally, the Higgs vacuum expectation value cannot vary by more than a factor of a few before atoms become unstable \cite{Agrawal:1997gf, Arkani-Hamed:2004fb}.  In the context of the string theory landscape populated by eternal inflation, there exists a natural setting for environmental selection to play out \cite{Susskind:2003kw,Nima:2005}.  This motivates considering models with parameters that may not be natural, but which are forced to be small by environmental selection pressure. 

Split supersymmetry is an example of a model that relaxes naturalness as a guiding principle and focuses on unification and dark matter \cite{Arkani-Hamed:2004fb, Giudice:2004tc,Wells:2003tf}.  In this theory, the scalar superpartners are ultra-heavy and the electroweak scale consists of one fine-tuned Higgs and the fermionic superpartners, which are kept light by R-symmetry.   The additional fermions alter the running of the gauge couplings so that unification occurs at high scales.

The ``minimal model''   was presented in \cite{Mahbubani:2006, Cirelli:2006}, where a Dirac electroweak doublet serves as a dark matter candidate and leads to gauge coupling unification.   A fermion singlet that mixes with the dark matter must also be introduced to avoid conflicts with direct detection results.  The additional fermion leads to richer phenomenology, but at the cost of introducing a new mass scale. 

In both these models, fermions serve as dark matter and technical naturalness protects their masses from large radiative corrections.  Split supersymmetry assumes that the high energy dynamics are supersymmetric, but that  high-scale susy breaking is preferred.  If the susy breaking sector communicates R-symmetry breaking inefficiently,  the gauginos and Higgsinos end up much lighter than the typical supersymmetric particles.  This should be contrasted with models like those in \cite{Mahbubani:2006}, where an ad hoc dynamical mechanism is invoked to make the fermionic dark matter much lighter than the GUT scale.    Without understanding how the ``minimal model'' fits into a high energy theory, it may be that it requires fine-tunings of fermion masses to get a viable dark matter particle.  Similarly, without understanding how R-symmetry breaking is explicitly communicated to the gauginos and Higgsinos, it may be that split susy requires a tuning of a fermion mass to get weak-scale dark matter.

The use of technical naturalness to justify new light fermions may be particularly misleading for relevant couplings that determine the large-scale structure of the Universe. 
 In \cite{Tegmark:2005dy,Hellerman:2005yi}, the formation of galactic structures with properties similar to the Milky Way places bounds on the ratio of the dark matter density to baryon density.  Typically, these bounds are not as strong as those on the cosmological constant, but they fix the dark matter mass to within an order of magnitude \cite{Tegmark:2005dy} (or three orders of magnitude in \cite{Hellerman:2005yi}). 
This opens up the possibility that the mass of the dark matter is unnatural and  is set by environmental conditions so that the baryonic fraction of matter is not  over- or under-diluted.   

A candidate for unnatural dark matter is the scalar WIMP.  While there has been some recent work on minimal models with scalar dark matter \cite{Pierce:2007, Barbieri:2006, Cirelli:2006, McDonald:1994,Davoudiasl:2005,Calmet:2006hs}, none provide a framework for gauge unification.  In this paper, we will study a minimal scalar unifon sector that also contains a viable dark matter candidate.  An additional Higgs doublet is added to the SM that is in a~{\bf 5} or~{\bf  6}-plet of a new global discrete symmetry.\footnote{Only the ~{\bf 5} option will be discussed here, but the results also hold for the~{\bf 6} multiplet.}  This global symmetry remains unbroken, yielding a spectrum of two five-plets of real neutral scalars and one five-plet of charged scalars.  Relic abundance calculations give the WIMP mass to either be $\sim 80$ GeV or in the range $\sim 200$-$700$ GeV.  The model will be tested by next-generation direct and indirect detection experiments, and may possibly have signatures at the LHC.  

The model will be presented in greater detail in Sec. II.  In Sec. III, the renormalization group equations are solved to illustrate gauge unification and the allowed weak-scale values of the theory's quartic couplings.  The relic abundance calculation is presented in Sec. IV and the predicted experimental signals, in Sec. V.  The results are summarized in Sec. VI.

\section{The Six Higgs Doublet Model} \label{model}

 The model proposed in this paper consists of the Standard Model Higgs, $h$, plus an additional electroweak doublet, $H_5$, in the {\bf 5} representation of a global discrete symmetry group.  The discrete symmetry is not necessary for maintaining the stability of the dark matter; its purpose is  to package the fine-tuning of the squared masses for the five additional doublets into a single tuning.  This discrete symmetry also reduces the number of quartic couplings in the potential to that of the two Higgs doublet model.  Any discrete symmetry group can be chosen, so long as it has a {\bf 5} representation (i.e., S$_6$).
  
The scalar potential for the six Higgs doublet model is
\begin{eqnarray}
V &=& - m_0^2 |h|^2+ m_5^2 |H_i|^2+\lambda_1 (|h|^2)^2+\lambda_2(|H_i|^2)^2 \nonumber\\
&&+\lambda_3 |h|^2|H_i|^2+\lambda_4 |h^{\dagger} H_i|^2+\lambda_5 ((h^{\dagger} H_i)^2+\mbox{h.c.}) \nonumber\\
&&+ \lambda_6 c_{ijk} (h H_i^{\dagger} H_j H_k^{\dagger}+\mbox{h.c.}),
\label{potential}
\end{eqnarray}
where $ i, j , k = 1, \ldots ,5$.  Depending on the choice of discrete symmetry, there may be several couplings of the form $|H_5|^4$; one possibility is shown in the $\lambda_2$ term.  The existence of the term proportional to $\lambda_5$ is necessary for the phenomenological viability of the model and forces the five-dimensional representation to be real.  The term proportional to $\lambda_6$ is only allowed if the symmetry satisfies the following relation
 \begin{equation}
{\bf 5} \otimes {\bf 5} = {\bf 1} \oplus {\bf 5} \oplus \cdots.
\end{equation}
The couplings $\lambda_5$ and $\lambda_6$ lead to a physical phase and will induce CP violation in the self-interaction of the  $\mathbf{5}$-plets after electroweak symmetry breaking.  This does not alter the tree-level spectrum; because the self-coupling of the {\bf 5}-plet is only affected at loop-level, the CP violation does not significantly alter the experimental signatures of the model.  

The field $h$ acquires a vev and gives masses to the gauge bosons.  In contrast, the field $H_5$ does not acquire a vev and cannot have any Yukawa interactions with the Standard Model fermions.  Expanding about the minimum of the potential, $\langle h\rangle = v/\sqrt{2}$, with $v = 246$ GeV.
The Higgs $\mathbf{5}$-plets are
\begin{eqnarray}
H_5=\left( \begin{array}{c}
\phi_5^+\\
(s^0_5+i a^0_5)/\sqrt{2}\end{array} \right). 
\end{eqnarray}
  The physical masses of the particles at the minimum are 
\begin{eqnarray}
m_{h^0}^2 &=& 2 \lambda_1 v^2 \nonumber \\ 
m_{\phi^\pm}^2 &=& m_5^2 +\frac{1}{2} \lambda_3 v^2 \nonumber \\
m_{s^0}^2 &=&  m_5^2 + \frac{1}{2} v^2 (\lambda_3+\lambda_4+2|\lambda_5|) \nonumber \\
m_{a^0}^2 &=&  m_5^2 + \frac{1}{2} v^2 (\lambda_3+\lambda_4-2|\lambda_5|) 
\end{eqnarray} 
and must always be greater than zero.  The lightest neutral particle, $a^0$, serves as the dark matter candidate; in order that it not become the charged $\phi^\pm$ boson, the quartics must satsify
\begin{equation}  \label{constraint1}
\lambda_4 - 2 |\lambda_5| < 0.
\end{equation}
The splitting between $s^0$ and $a^0$ is proportional to $\lambda_5$ and breaks the accidental $U(1)$ symmetry.  Results from direct detection experiments (see Sec. \ref{directdet}) require that the mass splitting between $s^0$ and $a^0$ be more than $\OO(100\keV)$, which sets the limit
\begin{equation}
|\lambda_5| \gsim 10^{-6}.
\end{equation}
The experimental lower bound on the Higgs mass \cite{Yao:2006} constrains the value of $\lambda_1$ to be 
\begin{equation}
\lambda_1 \gsim 0.1.
\end{equation}

Additional constraints on the quartics come from the requirement of vacuum stability.  In order that the potential (\ref{potential}) be bounded from below in all field directions, the couplings must satisfy
\begin{eqnarray} \label{constraint2}
\lambda_1, \lambda_2 &>& 0  \nonumber\\
\lambda_3&>&- 2 \sqrt{\lambda_1 \lambda_2} \nonumber\\
\lambda_3+\lambda_4 - 2 |\lambda_5| &>& - 2 \sqrt{\lambda_1 \lambda_2}.
\end{eqnarray}
These conditions are for local stability of the potential at a given scale.  If they are satisfied at all scales, then they correspond to absolute stability.

%If $h^{\dagger} h = x $ and $H_n^{\dagger} H_n = y $, then, by the Cauchy-Schwartz inequality, it follows that $h^{\dagger} H_n = c e^{i \alpha} \sqrt{xy}$ where $|c| \leq 1$.  The sum of the quartic terms in the scalar potential must by greater than 0  for all possible values of $x, y$, as well as the constants $c, \alpha$.  The first line of constraints in (\ref{constraint2}) comes from setting either $x$ or $y$ to 0.  The second line comes from setting $c=0$ (for all $\alpha$) and the third line results from $c=1$.  

\section{Renormalization Group Influence on Low Energy Spectrum} \label{LowEnergy}
\subsection{Gauge Unification}

Unification is the key motivation for introducing the five-plet $H_5$ and it is straightforward to check
that the gauge couplings unify reasonably well with the addition of six or seven scalars to the SM (see \cite{Machacek:1983gf} for two-loop RGEs) .  
%At one-loop, the renormalization group equations for the U(1), SU(2), and SU(3) gauge couplings are
%%
%\begin{equation} \label{oneloop}
 %\frac{d}{d t} \alpha_i^{-1} = -\frac{b_i}{2 \pi},
%\end{equation}
%%
%where $t=\log \mu$ and the $b_i$ are given in \cite{Machacek:1983gf} along with the complete two-loop beta functions.  It follows from (\ref{oneloop}) that unification holds when the following condition is satisfied:
%%
%\begin{equation} \label{approxrge}
%\frac{\alpha_1^{-1} - \alpha_2^{-1}}{\alpha_2^{-1} - \alpha_3^{-1}} \Bigg|_{\mu = m_{Z^0}} \simeq  \frac{b_1-b_2}{b_2-b_3}.
%\end{equation}
%%
% The r.h.s. of the expression is a function of the total number of scalar doublets, and leads to unification for seven. 
In particular, when two-loop RGEs are evaluated, a threshold correction splitting a (fermionic and scalar) $\mathbf{5} + \mathbf{\overline{5}}$  by $m_2/m_3 \simeq 30$ is necessary to maintain unification for the six scalar case (where $m_2$ and $m_3$ are, respectively, the masses of a doublet and triplet).  This is better than the case of seven scalars, where $m_2/m_3 \simeq 300$.  As a point of comparison, the threshold corrections for the MSSM require $m_3/m_2 \simeq 20$ \cite{Murayama:2001ur}.    

%\begin{figure}[t]   
%\includegraphics[width=3 in]{RGE.pdf}
%\caption{Two-loop running of the SM gauge couplings for $N_s = 6$ (top) and $N_s=7$ (bottom).  In both plots, the couplings unify at $M_\GUT \simeq 10^{14}$ GeV.  The experimental value of $\alpha_s(m_z)$ is 0.1176$\pm$0.0020 \cite{Yao:2006}.}
%\label{RGE:GaugeCouplings}
%\end{figure}

The unification scale is given by
\begin{eqnarray}
t_\GUT =  2\pi \frac{ \alpha_1^{-1} - \alpha_2^{-1}}{ b_1 - b_2} \Rightarrow M_{\text{GUT}} \simeq 10^{14}\GeV
\end{eqnarray}
and the value of the gauge coupling at the GUT scale is
\begin{eqnarray}
\alpha_{\text{GUT}}^{-1} = \alpha_2^{-1} -\frac{ b_2}{2\pi} t_\GUT\simeq 40.
\end{eqnarray}
If this theory is embedded in a simple $SU(5)$ GUT, the resulting six-dimensional proton decay 
is
\begin{eqnarray}
\Gamma(p\rightarrow e^+ \pi^0) \simeq \frac{\alpha_{\text{GUT}}^2 m_p^5}{M_{\text{GUT}}^4} \simeq 10^{-35} \mbox{ s$^{-1}$}, 
\end{eqnarray}
which is far too fast.  This implies that GUT-scale physics is non-minimal and must suppress gauge-mediated proton decay.  Several approaches exist in the literature to deal with this task.  One possibility, discussed in \cite{Mahbubani:2006}, is to embed the theory in a five-dimensional orbifold model.  Proton decay is still allowed, but is highly suppressed due to the configuration of the fields in the extra dimensions.  Trinification, a GUT  based upon the group $[SU(3)]^3$, provides another option because it completely forbids proton decay via gauge bosons \cite{Glashow:1984gc}.  

\subsection{Quartic Couplings}

The ability to discover the six Higgs dark matter candidate depends on its mass and its couplings to SM particles.  The gauge interactions are fixed, but the couplings to the SM Higgs are model-dependent.  These couplings must satisfy two requirements: perturbativity and vacuum stability.  In addition, they must adhere to experimental constraints from Higgs and dark matter searches (see Sec. \ref{model}).  The model dependence comes into play when choosing the value of the quartics at the GUT scale.  

The most common understanding of how fine-tuning can give rise to Higgs and dark matter candidates near the electroweak scale invokes a landscape of vacua, each with its own values for couplings and masses.  The string theory landscape allows for a great range of possibilities for the physical parameters of the theory and naturally leads to the question of what the typical values are in our neighborhood of vacua.   The distribution of couplings is clearly a UV sensitive question and cannot be obtained by dimensional analysis because the quartics are dimensionless.  Fortunately, there are simple ansatze that lead to distinct weak-scale spectra.   

This section will explore two possible GUT-scale distributions of the quartics: parameter space democracy and susy.  The couplings at the weak scale are obtained by application of the renormalization group equations.  The resulting differences in weak-scale phenomenology for each of these distributions will be explored in Sec.~\ref{ExperimentalSignatures}.

\subsubsection{Parameter Space Democracy}
Perhaps the most obvious distribution of parameters is one where all couplings at the GUT scale are equally probable -- ``parameter democracy.''  This measure favors large couplings of either sign. If all the couplings are positive, they quickly run down to perturbative values and the initial boundary conditions of the quartics are not terribly important.  When the quartic couplings start off negative, they can become asymptotically free and may have Landau poles.  Furthermore, negative quartic couplings can lead to vacuum decay, especially when they have a large magnitude initially; thus, most of the parameter space in the negative direction is ruled out.

Typically, the couplings approach a tracking solution rather rapidly \cite{Pendleton:1980as}.  Neither $\lambda_5$ nor $\lambda_6$ has a significant affect on the fixed point values of the other couplings.  The gauge boson and top quark contributions also do not significantly affect the runnings in this region.  With these observations, the beta functions may be approximated as 
\begin{equation}  \label{lambdaRGE}
16 \pi^2 \frac{d \lambda_i}{d t} = b_{\lambda_i},
\end{equation}
where
\begin{eqnarray}
b_{\lambda_1} &\simeq& 24 \lambda_1^2 + 2 N_h \lambda_3^2 + 2 N_h \lambda_3 \lambda_4 + N_h \lambda_4^2\nonumber \\
b_{\lambda_2}&\simeq& 4 (2 N_h+4) \lambda_2^2 + 2\lambda_3^2 + 2 \lambda_3 \lambda_4 \nonumber\\
b_{\lambda_3} &\simeq& 4 \lambda_3^2+2 \lambda_4^2 + 4 \lambda_4 (\lambda_1+ N_h \lambda_2)\nonumber\\ &&+ 4 \lambda_3 (3 \lambda_1+(2 N_h +1) \lambda_2) \nonumber\\ 
b_{\lambda_4} &\simeq& 4 \lambda_4 (\lambda_1+ \lambda_2) + 4 \lambda_4^2+ 8 \lambda_3 \lambda_4 \nonumber
\end{eqnarray}
and $N_h$ is the number of scalars added to the SM, in addition to the usual Higgs \cite{Inoue:1980}.   For the model considered here, $N_h = 5$.  

\begin{figure}[t] 
\includegraphics[width=3.3 in]{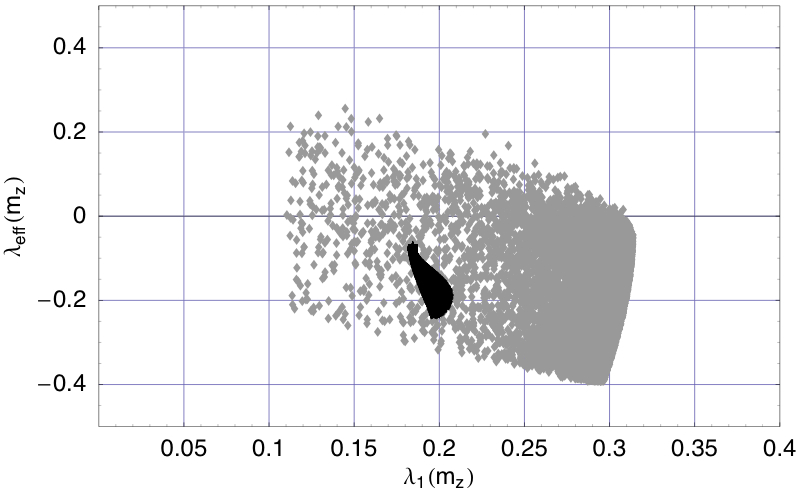}
\includegraphics[width=3.3 in]{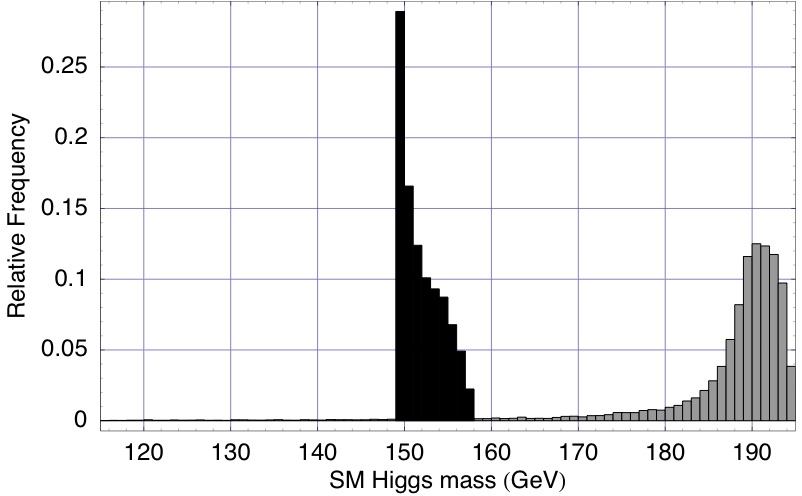}
\caption{(Top) Distribution of $\lambda_1$ and $\lambda_{\eff} = \lambda_3 + \lambda_4 - 2 |\lambda_5|$ at the electroweak scale obtained by solving the one-loop renormalization group equations for parameter space democracy (gray) and susy (black) boundary conditions at the GUT-scale.  The couplings are focused down to a small range at the weak scale.  (Bottom)  The distribution of SM Higgs mass for the two sets of boundary conditions.  The distribution of allowed Higgs masses is smaller in the case of susy boundary conditions as opposed to parameter space democracy conditions. }
\label{Dist:Couplings}
\end{figure}

When $\lambda_1$ and $\lambda_2$ are large at the GUT scale, the self-coupling terms in the beta functions (\ref{lambdaRGE}) dominate and the low energy values for these couplings are approximately
\begin{equation} \label{lambdas}
\lambda_1^\Max \simeq \frac{16 \pi^2}{24t_\GUT} \sim 0.24 \quad \quad \lambda_2^\Max \simeq \frac{16 \pi^2}{120t_\GUT} \sim 0.05. 
\end{equation}
Figure~\ref{Dist:Couplings} (gray points) shows the weak-scale distribution of $\lambda_1$ and the effective coupling
\begin{equation}
\lambda_{\eff} = \lambda_3 + \lambda_4 - 2 |\lambda_5|,
\end{equation}
which parametrizes the interaction of the WIMP candidate $a_5^0$ to the SM Higgs $h^0$.  The values of the quartics at the GUT scale were randomly sampled within the range:  $0 \lesssim \lambda_1, \lambda_2, |\lambda_3| \lesssim \OO(4 \pi)$, $-1 \lesssim \lambda_4 \lesssim 0$, and $|\lambda_5| \lesssim 2$.\footnote{Quartics outside this range either give the same result for the low-energy spectra or, as in the case of $\lambda_5$, cause the couplings to run down to non-perturbative values.  This range was chosen to maximize the sampling rate of the program.}  They were then run down to the electroweak scale by applying the renormalization group equations.  Despite the large range of possibilities at UV energies,  the couplings are focused down to a narrow set at electroweak energies.  Indeed, $\lambda_1$ and $\lambda_2$ do not vary much from the values approximated in (\ref{lambdas}).  The region of parameter space at the electroweak scale corresponds to 
\begin{eqnarray}
0.1\lsim \lambda_1 \lsim 0.3,\quad
0\lsim \lambda_2 \lsim 0.1, \nonumber\\
-0.2 \lsim \lambda_3 \lsim 0.4,\quad
-0.5 \lsim \lambda_4 \lsim 0.
\end{eqnarray}
$\lambda_5$ renormalizes itself and thus remains small ($|\lambda_5| \lsim 0.1$).   With the assumption of parameter democracy,  the model comes close to saturating the upper values for $\lambda_1$ and $\lambda_2$, having a small $\lambda_3$, and having a $\lambda_4$ that is close to saturating the lower bound.    The acceptable range for the Higgs mass is 
\begin{eqnarray}
114\GeV \lesssim m_{h^0} \lesssim 200\GeV.
\end{eqnarray}
Higgs masses at the upper-end of this interval are preferred (see Fig.~\ref{Dist:Couplings}). 

%The running of  $\lambda_3$ depends more intricately on $\lambda_1$ and $\lambda_2$.  When $\lambda_3$ is on the order of these two couplings, it reaches a tracking solution between $\lambda_1^\Max$ and $\lambda_2^\Max$.  When it is larger than this value, it runs down to a value greater than $\lambda_1^\Max$ and starts to `push' both $\lambda_1$ and $\lambda_2$ down in value, potentially driving them negative, which leads to vacuum decay problems.  In general, $\lambda_4$ holds up all the other couplings; when $\lambda_4$ is negative, the other couplings are guaranteed to be positive.  As $\lambda_4$ becomes positive, the other couplings may become negative, which is particularly bad in the case of $\lambda_1$ and $\lambda_2$.  

\subsubsection{Minimal Susy Boundary Conditions}

Another plausible set of boundary conditions are ones where supersymmetry is broken at the GUT scale and the dominant quartic couplings are those arising from D-terms.   The simplest way of achieving the desired low-energy spectrum is if each low-energy Higgs doublet comes from a vector-like chiral superfield:
$\Phi_h$ and $\Phi^c_h$ for the Standard Model Higgs and $\Phi_{H_5}$ and $\Phi^c_{H_5}$ for the five-plet of scalar dark matter.  Specifically,   
\begin{eqnarray}
\nonumber
\Phi_h| = c_\beta  h - s_\beta \tilde{h}&\quad& \Phi_h^c| = s_\beta  h^\dagger + c_\beta \tilde{h}^\dagger\\
\nonumber
\Phi_{H_5}|= c_{ \beta_5}  H_5 - s_{\beta_5} \tilde{H_5}&\quad& \Phi_{H_5}^c|= s_{\beta_5}  H_5^\dagger + c_{\beta_5} \tilde{H}_{5}^\dagger,
\end{eqnarray}
where $\beta$ and $\beta_5$ are the orientation of the scalars inside the chiral superfields.
The resulting $D$-term potential has the following couplings
\begin{eqnarray}
\nonumber
\lambda_1 = \frac{2}{5} g^2_\GUT c^2_{2\beta}&\quad&
\lambda_2 =\frac{2}{5} g^2_\GUT c^2_{2\beta_5}\\
\nonumber
\lambda_3=- \frac{7}{10} g^2_\GUT c_{ 2\beta}c_{ 2 \beta_5}&\quad&
\lambda_4=g^2_\GUT c_{ 2\beta}c_{ 2 \beta_5}\\
\lambda_5=0&\quad&\lambda_6=0.
\end{eqnarray}
A term must be added to the superpotential to generate $\lambda_5 = 0$.  In order that this not alter the above relations significantly, it should be a small coupling.  One possibility is to take the minimum value allowed by direct detection experiments, $\lambda_5 \sim 10^{-6}$.  The gauge couplings at the GUT scale are $g_\GUT^2=0.32$, so the susy boundary conditions result in small couplings at the electroweak scale.  In order to have neutral dark matter, $\lambda_4<0$, so $\cos 2 \beta \cos 2\beta_5<0$. 

These couplings are  a function of two angles and lead to a lighter Higgs and smaller mass splittings for the scalars than the case of parameter space democracy.  This is apparent from Figure 1, where the black points show the allowed values of $\lambda_1$ and $\lambda_{\eff}$ at the electroweak scale obtained using the susy boundary conditions.  In this case, the Higgs mass falls within a much smaller range
\begin{equation}
147\GeV \lesssim m_{h^0} \lesssim 159\GeV
\end{equation}
and is lighter than the most probable Higgs mass for parameter space democracy.  

%The sharp peak in the relative frequency of the Higgs mass around $\sim$ 149 GeV can be explained from the beta function for $\lambda_1$; it follows from the limit of small quartic couplings, when the only terms that dominate the beta function are those that depend on the gauge and top Yukawa couplings.    

\section{Dark Matter} \label{DarkMatter}

\subsection{Relic abundance} \label{relic}

The lightest neutral component of the $H_5$ doublet, $a_5^0$, is a viable candidate for the observed dark matter and its mass may be estimated from standard relic abundance calculations.  It is assumed that the $a_5^0$ is in thermal equilibrium during the early universe.  When the annihilation rate of the $a_5^0$ is on the order of the Hubble constant, its number density `freezes out,' resulting in the abundance seen today.  

When the freeze-out temperature is on the order of the mass splittings $\Delta m_{s^0a^0}$ and $\Delta m_{\phi^{\pm}a^0}$, the presence of the additional scalars $s_5^0$ and $\phi^+_5$ becomes relevant \cite{Griest:1991}.  In this case, interactions involving the two other scalars as initial state particles are important in  determining the relic abundance of $a_5^0$, which must fall within the WMAP region $0.099 < \Omega_{\text{dm}} h^2 < 0.113$, where $\Omega_{\text{dm}}$ is the dark matter fraction of the critical density and $h = 0.72 \pm 0.05$ is the Hubble constant in units of 100 km s$^{-1}$ Mpc$^{-1}$) \cite{Spergel:2007}.   Typically, coannihilation has a significant effect on the allowed mass range of the relic.   

The number density of $a_5^0$ is given by
\begin{equation} \label{Boltzmann}
\frac{d n}{dt} = -3 H n - \sum_{i,j = a^0,s^0,\phi^{\pm}} \langle \sigma_{ij} v_{ij} \rangle (n_i n_j - n_i^{eq} n_j^{eq}),
\end{equation}
where $\sigma_{ij}$ is the sum of the annihilation cross sections of the new scalars $X_i$ into Standard Model particles $X$
\begin{equation}
\sigma_{ij}=  \sum_{X} \sigma(X_i X_j \rightarrow X X). 
\end{equation}
The first term on the r.h.s. of equation (\ref{Boltzmann}) accounts for the decrease in the relic density due to the expansion of the universe; the second term results from dilution of the relic from interactions with other particles.  The annihilation rate depends on the number of scalars added to the theory in addition to the SM Higgs, $N_h$, through the interaction cross sections $\sigma_{ij}$.  In general, $\sigma_{ij} \propto N_h^{-1} m_a^{-2}$, so the mass of the dark matter scales as 
\begin{equation}  \label{mass}
m_{a^0} \propto \frac{1}{\sqrt{N_h}}.
\end{equation}
Thus, in the non-resonance regime, the dark matter mass decreases with the number of electroweak doublets added to the SM.  For this reason, the six Higgs doublet model gives lighter dark matter than the inert doublet model \cite{Barbieri:2006}. 

When $m_{a^0} \lesssim 80$ GeV, the only annihilation channel is to a pair of fermions.  Because these cross sections tend to be rather small, $\Omega_{\text{dm}} h^2 \gtrsim 0.1$.  However, a resonance due to $s$-channel SM Higgs exchange causes a sharp decrease in the relic density $\sim80$ GeV, bringing it within the WMAP experimental range.  For $m_{a^0} \gtrsim 80$ GeV, diboson production is the dominant annihilation mechanism and keeps the abundance small.  There is always another point in this large mass regime where $\Omega_{\text{dm}} h^2 \sim 0.1$.  Thus, the dark matter can take two possible mass values - one light ($\sim 80$ GeV) and the other heavy ($\gtrsim 200$ GeV).   

\begin{figure}[t] 
\includegraphics[width=3.5in]{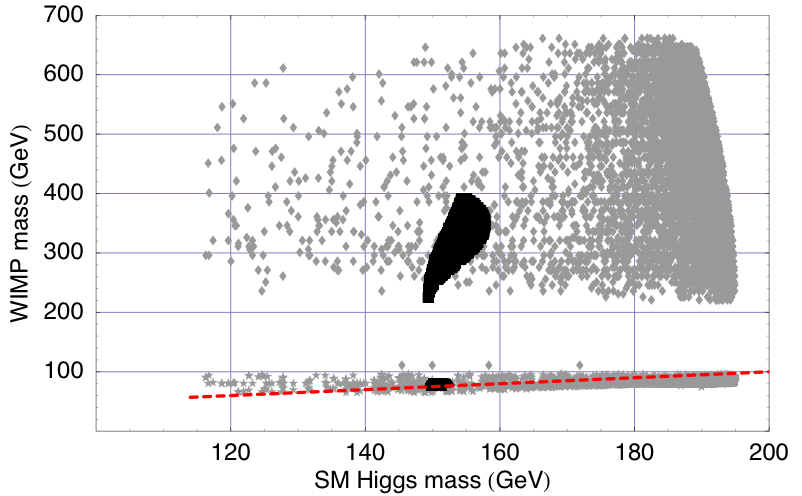}
\caption{
Allowed mass of the LSP $a_5^0$ as a function of the Higgs mass for parameter space democracy (gray) and susy (black) boundary conditions.  All points included in this plot fall within 1$\sigma$ of the electroweak precision data (Sect.~\ref{ewptsection}) and are consistent with LEP results (Sect.~\ref{lepresults}).  The SM Higgs can decay into a pair of WIMPs if $m_{a^0}$ lies below the dashed red line. }  
\label{DarkMatter:Mass}
\end{figure}

The relic abundance calculation was performed numerically by scanning over the parameter $m_{a^0}^2$ for each set of randomly selected quartic couplings.  Figure~\ref{DarkMatter:Mass} is a plot of the allowed mass of $a_5^0$ as a function of the SM Higgs mass.  A broad range of values is allowed for the case of parameter space democracy, with $m_{a^0}$ falling between $\sim 200 - 700$ GeV.  For supersymmetric boundary conditions, the mass values range from $\sim 200-400$ GeV.  An $\sim 80$ GeV dark matter particle is also allowed for both cases.  

\subsection{Bounds from electroweak precision tests} \label{ewptsection}

Electroweak precision tests place limits on the light mass range of the dark matter \cite{Peskin:1991sw}.  %The T parameter quantifies the strength of radiative corrections to the electroweak parameter $\rho$ and can be calculated most easily using the wavefunction renormalization of the Goldstone bosons $\omega^+$ and $\omega^0$ \cite{Barbieri:1993}: 
%%
%\begin{equation}
%\Delta \rho = \alpha T = \delta Z_{\omega^\pm}-\delta Z_{\omega^0}
%\end{equation}
%%
%The scalar potential (\ref{potential}) must be expanded in terms of $\phi^+$ and $\chi$; the relevant terms are
%%
%\begin{eqnarray}
%V &\supset& \frac{v}{\sqrt{2}} \omega^\pm \phi^\mp \Big[ s^0 (\lambda_4 + 2 \lambda_5) + i a^0 (\lambda_4 - 2 \lambda_5) \Big]  \nonumber \\ &&+ 2 \sqrt{2} v \lambda_5 a^0 s^0 \omega^0
%\end{eqnarray}
%%
The contribution of the new particles to the T parameter is given by 
\begin{eqnarray}
\Delta T &=& \frac{N_h}{16 \pi^2 \alpha v^2} \Big[F(m_{\phi^\pm}, m_{a^0})+ F(m_{\phi^\pm}, m_{s^0}) \nonumber \\
&& - F(m_{a^0}, m_{s^0})\Big] ,
\end{eqnarray}
where 
\begin{eqnarray}
F(m_1, m_2) = \frac{m_1^2+m_2^2}{2}-\frac{m_1^2 m_2^2}{m_1^2-m_2^2} \log\frac{m_1^2}{m_2^2}.
\end{eqnarray}
The expression for $F(m_1, m_2)$ can be simplified if one assumes that the mass splitting $\Delta m = m_2 - m_1$ satisfies $\Delta m/ m_1 \ll 1$.  In this limit,
\begin{equation}
F(m, m+\Delta m) = \frac{2}{3} (\Delta m)^2 + \OO\left(\frac{(\Delta m)^4}{m^2}\right)
\end{equation}
and the expression for $\Delta T$ reduces to     
\begin{eqnarray} \label{deltaT}
\Delta T &\simeq& \frac{N_h}{12 \pi^2 \alpha v^2} (m_{\phi^\pm}-m_{a^0}) (m_{\phi^\pm}-m_{s^0}) \nonumber \\
&\simeq& \frac{ N_h v^2}{192 \pi^2 \alpha m_a m_s} (\lambda_4^2 -4 \lambda_5^2).
\end{eqnarray}

Because $\lambda_5$ is typically smaller than $\lambda_4$, $\Delta T$ is always positive.  In the minimal SM, $\Delta T$ is driven more negative as the mass of the Higgs increases.  The additional scalar doublet $H_5$ compensates for this change, driving $T$  positive.  

The $S$ parameter also has contributions from the additional Higgs doublets  \cite{Barbieri:2006} and is easily generalized to the case of six Higgses
\begin{equation}
\Delta S = \frac{N_h}{2 \pi} \int_0^1 x (1-x) \log\Big[\frac{x m_{s^0}^2+(1-x) m_{a^0}^2}{m_{\phi^{\pm}}}\Big] dx.
\end{equation}
When the mass splittings are small,
\begin{eqnarray}
\Delta S &=& \frac{N_h}{12 \pi m_{a^0}} \Big(\Delta m_{s^0a^0} - 2 \Delta m_{\phi^\pm a^0} \Big)  + \mathcal{O}\left(\frac{\Delta m^2}{m}\right) \nonumber \\
&\simeq& \frac{N_h v^2 \lambda_4}{24 \pi m_{a^0}^2}.
\end{eqnarray}

For the heavy dark matter candidate, the corrections to the $S$ and $T$ parameters fall well within the 1$\sigma$ electroweak precision data \cite{Barbieri:2006}.   Lighter dark matter can make significant contributions to the $S$ and $T$ parameters.  Couplings that give rise to deviations in $S$ and $T$ that are more than 1$\sigma$ away from the measured values have not been used in the analysis of the experimental signatures of the model (Sect.~\ref{ExperimentalSignatures}).      

%\begin{figure}[t]
%\includegraphics[width=3.3 in]{EWPT.pdf}
%\caption{Corrections to the electroweak S and T parameters in the six Higgs doublet model for parameter space democracy (grey) and susy (black) boundary conditions.  Points corresponding to the heavy dark matter are encircled by the dotted line.  The red line corresponds to the 1$\sigma$ experimental bound.}
%\label{ewpt}
%\end{figure}

\section{Experimental Signatures} \label{ExperimentalSignatures}

\subsection{Direct detection} \label{directdet}

Direct detection experiments provide a means for observing the dark matter relic when it scatters elastically off atomic nuclei \cite{Jungman:1996}.  The WIMP can either couple to the spin of the nucleus or to its mass.  The spin-independent contribution to the cross section usually dominates and bounds on its value are being set by experiments such as CDMS, DAMA, Edelweiss, ZEPLIN-I, and CRESST.

\begin{figure}[t]
\includegraphics[width=3.3 in]{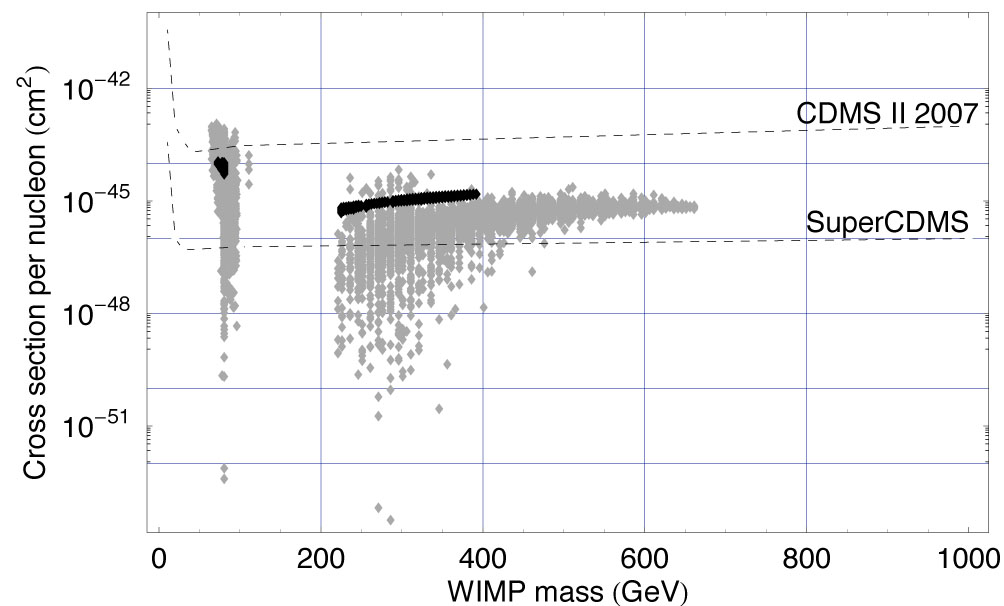}
\caption{Cross section per nucleon for the case of parameter space democracy (gray) and susy (black) boundary conditions.  The lightest WIMP candidates will be tested for at the current CDMS II run (upper dashed line).  The third phase of SuperCDMS (lower dashed line) will probe a greater region of the parameter space.  }
\label{DirectDetection}
\end{figure}

In the six Higgs doublet model, there are two contributions to the spin-independent cross section.  The first comes from an $s$-channel Higgs exchange described by the effective Lagrangian 
\begin{equation}
L_{\eff} = \sum_q \left(\frac{-i \lambda_\eff}{m_{h^0}^2}\right) m_q a^0 a^0 q \bar{q}.
\end{equation}
%The cross section for Higgs exchange between the relic and nucleus is \cite{Jungman:1996, Kitano:2005ew,Barbieri:2006}
%%
%\begin{equation}
%\sigma = \frac{\mu^2}{4 \pi m_{a^0}^2} \left(\frac{\lambda_\eff^2} { m_{h^0}^4} \right) \cdot [Z f_p + (A-Z) f_n]^2
%\end{equation}
%%
%where $\mu = m_{a^0} m_N/(m_{a^0} + m_N)$ is the reduced mass of the WIMP-nucleus system, $Z$ is the proton number, $A$ is the atomic number, and $f_p \simeq f_n$ is the matrix element 
%%
%\begin{equation}
%\langle p | \sum _q m_q q \bar{q} | p \rangle \simeq 0.3 m_p,
%\end{equation}
%%
%with $m_p$ the proton mass.  
Experimental results are usually reported in terms of the cross section per nucleon, which in this case is  
\begin{equation}
\sigma_n = 2 \times 10^{-9}\text{pb} \left(\frac{\lambda_{\text{eff}}}{0.4}\right)^2 \left(\frac{350\GeV}{m_{a^0}}\right)^2 \left(\frac{200\GeV}{m_{h^0}}\right)^4. 
\end{equation}
This cross section scales as $N_h$ (see Eq. \ref{mass}).  Because $\sigma_n \propto m_{a^0}^{-2}$, the lighter dark matter candidate will have a stronger signal than its heavier counterpart.  Figure~\ref{DirectDetection} shows the cross section per nucleon for the case of parameter space democracy (gray) and susy (black) boundary conditions.  The current CDMS II run is sensitive to the lightest WIMPs predicted by the model.  A larger portion of the parameter space is within the testable reach of the proposed SuperCDMS experiment \cite{Akerib:2006}.  The lower dashed line on the plot is the expected limit from Phase C of SuperCDMS.       

Another contribution to the spin-independent cross section comes from the inelastic vector-like interaction $a^0 + p \rightarrow s^0 + p$, which is mediated by an off-shell $Z^0$-boson.  In general, such inelastic transitions provide a means to reconcile DAMA's detection of relic-nucleon scattering, which conflicts with CDMS's null result \cite{Bernabei:2003}.  Consistency with the experimental results requires a mass splitting $\Delta m_{s^0a^0} \simeq 100$ keV between the two lightest scalars \cite{Smith:2002af}. 

\subsection{Indirect detection}

A concentration of WIMPs in the galactic halo increases the probability that they will annihilate to produce high-energy gamma rays and positrons \cite{Bergstrom:1997}.  The gamma ray signal is of particular interest because it is not scattered by the intergalactic medium; thus, it should be possible to extract information about the WIMP mass from the spectrum.  

Monochromatic photons can be produced when the WIMP annihilates to produce $\gamma \gamma$ and $Z^0 \gamma$.  The dominant mechanisms that contribute to this annihilation depend on the DM mass regime.  The light dark matter, for example, annihilates primarily through $s$-channel Higgs exchange with a one-loop $h^0\gamma X$ vertex (X = $\gamma$, Z$^0$).  The main contributions to the loop come from the $W^\pm$ boson, the top quark, and the $\phi_5^\pm$ five-plet.  Other box diagrams are suppressed.  The WIMPs are highly non-relativistic and their annihilation cross section in the light mass regime is nearly
\begin{equation}
\sigma(a^0 a^0 \rightarrow \gamma X) u \simeq \frac{1}{N_h} \frac{v^2 \lambda_\eff^2}{(s-m_{h^0}^2)^2+m_{h^0}^2 \Gamma_{h^0}^2} \frac{\Gamma({h^0} \rightarrow \gamma X)}{\sqrt{s}},
\end{equation}  
where $u$ is the relative velocity between the initial two WIMPs and $s  \approx 4 m_{a^0}^2$.  The general expressions for the decay widths of the Higgs boson into a $\gamma \gamma$ and $\gamma Z^0$ final state are found in \cite{HiggsHunter, Birkedal:2006}. 

The case of the heavy dark matter is significantly different \cite{Hisano:2005}.  In this regime, the dominant contribution comes from the box diagram with three $\phi_5^{\pm}$ and one $W^+$ in the loop.  When the $a_5^0$ and $\phi_5^{\pm}$ are nearly degenerate and $m_{a^0} \gg m_{W^\pm}$,  there is an effective long-range Yukawa force between the $\phi_5^+ \phi_5^-$ pair in the loop that is mediated by the gauge boson:
\begin{equation}
V(r) \sim - \alpha_2 \frac{e^{-m_{W^\pm} r}}{r}.
\end{equation}

 As a result, the pair of charged scalars form a bound-state solution to the non-relativistic Schrodinger equation.  The optical theorem is used to obtain the s-wave production cross section for the bound state:
\begin{equation}
\sigma(a_5 a_5 \rightarrow \phi_5^+ \phi_5^-) u \sim \frac{2 \alpha_2^2 m_{a^0}^2}{N_h m_{W^\pm}^2} \Bigg(1+\sqrt{\frac{2 m \Delta m_{\phi^{\pm} a^0}}{m_{W^\pm}^2}}\Bigg)^{-2}.
\end{equation}
Multiplying this by the decay width of the bound state to two photons (or, $\gamma Z^0$), gives the total annihilation cross section
\begin{equation}
\sigma(a_5 a_5 \rightarrow \gamma \gamma) u \sim \frac{2 \pi \alpha^2 \alpha_2^2}{N_h m_{W^\pm}^2} \Bigg(1+\sqrt{\frac{2 m \Delta m_{\phi^{\pm} a^0}}{m_{W^\pm}^2}}\Bigg)^{-2}.
\end{equation} 
This cross section does not depend on $m_{a^0}$ (to zero-th order in the mass splittings) and, as a result, is significantly enhanced in the heavy DM mass region.  This enhancement is critical; because of it, the heavy mass DM may be visible in gamma ray experiments.  Additionally, the only parameter dependence comes in through the mass-splittings, which are small.  Therefore, there is not much spread in the range of allowed cross sections. 

\begin{figure}[b]
\includegraphics[width=3.3 in]{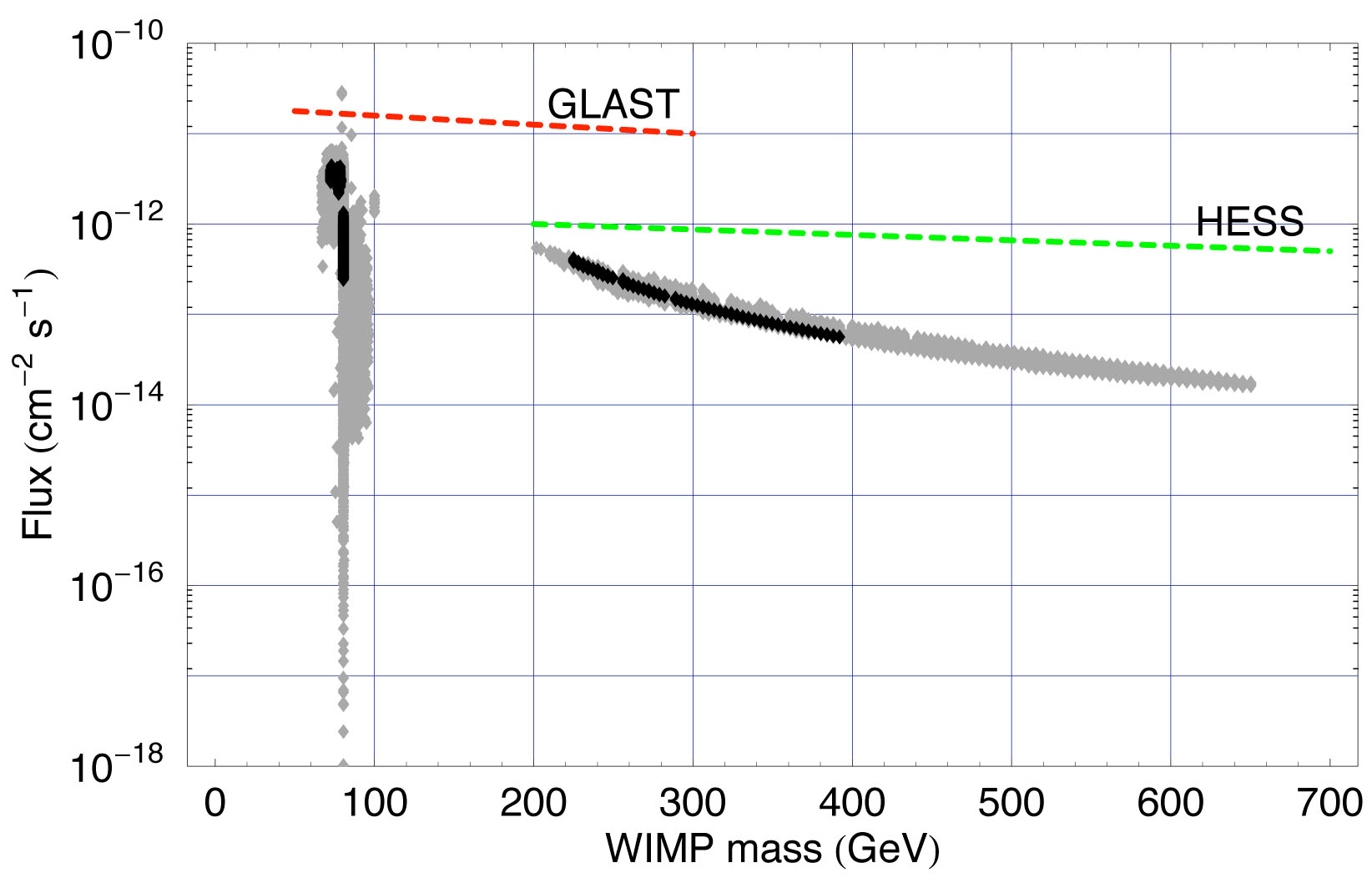}
\caption{Approximate flux from dark matter annihilation in the galactic halo via $a_5^0 a_5^0 \rightarrow \gamma \gamma$ for parameter space democracy (gray) and susy (black) boundary conditions.  The dashed lines indicate the sensitivity of GLAST (red) and the ground-based detector HESS (green).  The NFW profile was used.}
\label{IndirectDetection}
\end{figure}

%\begin{figure}[b]
%\includegraphics[width=3.3 in]{LEPplot.pdf}
%\caption{The dotted line is the production cross section for $e^+ e^- \rightarrow \phi^+ \phi^-$ at LEP as a function of $m_{\phi^{\pm}}$.  The solid line is the production cross section for $e^+ e^- \rightarrow a_5^0 s_5^0$ as a function of $m_{s^0}$ (GeV), for the case $m_{a^0} = 80$.  The shaded regions are excluded by LEP \cite{Abbiendi:2004}. \\              }
%\label{LEP}
%\end{figure}

The monochromatic flux due to the gamma ray final states observed by a telescope with a field of view $\Delta \Omega$ and line of sight parametrized by $\Psi = (\theta, \phi)$ is given by
\begin{equation}
\Phi = C_{\gamma X} \left(\frac{\sigma_{\gamma X} u}{1 \mbox{ pb}}\right) \Big(\frac{100 \mbox{ GeV}}{m_{a^0}}\Big)^2 \bar{J}(\Psi, \Delta \Omega) \Delta\Omega,
\end{equation}
where
\begin{eqnarray}
\nonumber
C_{\gamma \gamma} &=&1.1 \times 10^{-9} \text{cm}^{-2}\text{s}^{-1}\nonumber \\
C_{\gamma Z^0}&=&5.5 \times 10^{-10}  \text{cm}^{-2}\text{s}^{-1}
\end{eqnarray}
and the function $\bar{J}$ includes the information about the dark matter distribution in the halo.  Note that the flux is independent of $N_h$.  For the NFW profile, $\bar{J} \simeq 10^3$ for $\Delta \Omega = 10^{-3}$ \cite{Bergstrom:1997}.  Other profile models exist with either more mildly/strongly cusped profiles at the galactic center \cite{Birkedal:2006}.  Depending on which model is chosen, $\bar{J}$ can be as small as $10$ or as large as $10^5$ for $\Delta \Omega = 10^{-3}$.  In this work, the moderate NFW profile will be used, however the result can easily be scaled by two orders of magnitude to get the predictions for other halo profiles.  
%%
%\begin{equation}
%\bar{J}(\Psi, \Delta \Omega) \equiv \frac{1}{8.4 \mbox{ kpc}} \left(\frac{1}{0.3 \mbox{GeV/cm$^3$}}\right)^2 \frac{1}{\Delta \Omega} \int_{\Delta \Omega} \!\!\!\!d\Omega \int_{\Psi}\!\! \rho^2 dl
%\end{equation}
%%

The expected monochromatic flux for the $\gamma \gamma$ line is shown in Figure~\ref{IndirectDetection} (assuming $\bar{J} \Delta \Omega =1$).  The estimated flux is right beneath the sensitivity limits of the ground-based HESS detector (green line) and space-based GLAST telescope (red line) \cite{Birkedal:2006, Zaharijas:2006}.  The results for the $\gamma Z^0$ line are similar for low DM masses and are enhanced by an order of magnitude for masses greater than 200 GeV, putting it within the reach of HESS.  Given the two order of magnitude uncertainty in the flux coming from the details of the halo profile, the gamma ray line is an interesting signal for both current and upcoming experiments.

\subsection{Collider Signatures}  \label{lepresults}

\begin{figure}[b]
\includegraphics[width=3.3 in]{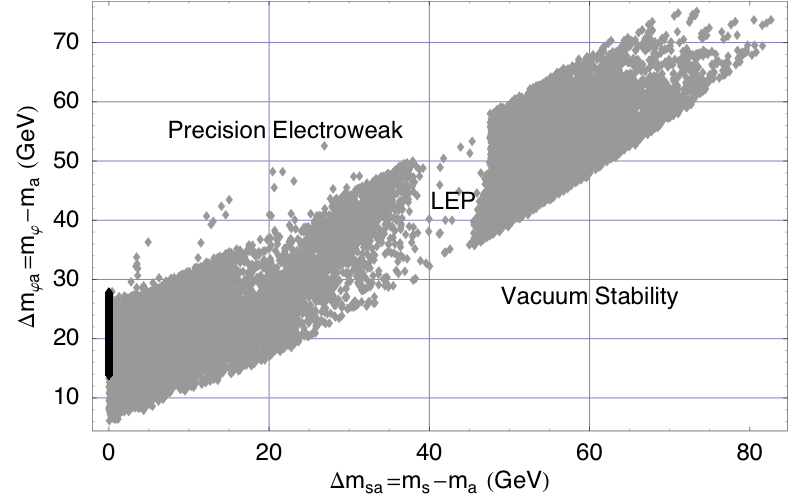}
\caption{Mass splittings for the light dark matter ($m_{a^0} \sim 80$ GeV).  LEP excludes the region of intermediate $s_5$ mass.  The region on the upper left is excluded by electroweak precision results, while that on the lower right is excluded by  vacuum stability. Results are shown for parameter space democracy (gray) and susy (black) boundary conditions.}
\label{MassSplittings}
\end{figure}

\begin{figure}[b]
\includegraphics[width=3.3 in]{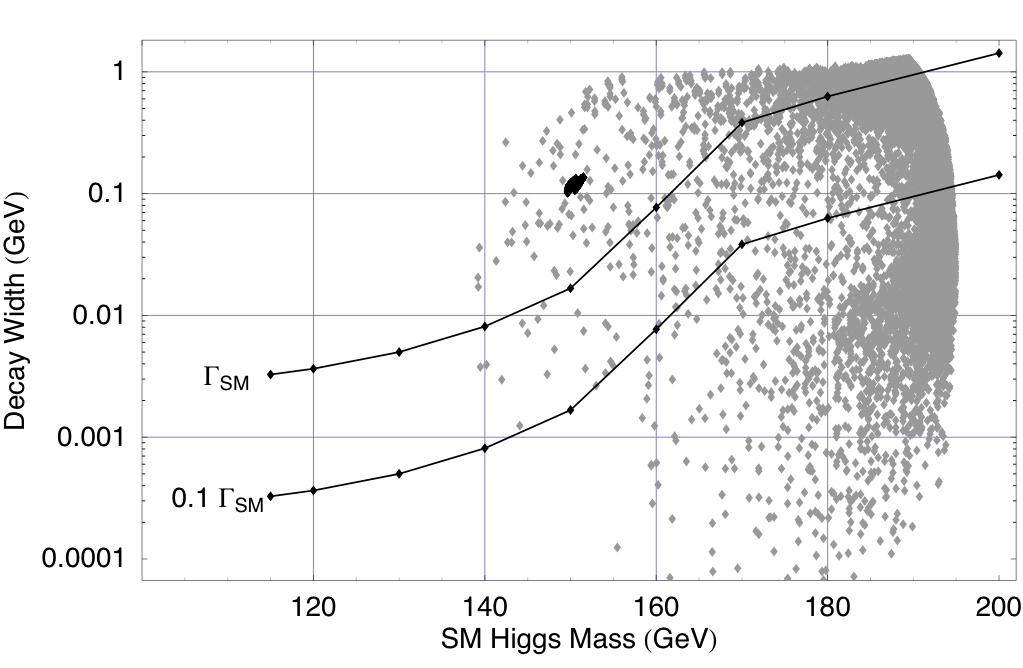}
\caption{Width of the SM Higgs decay into the $H_5$ scalars for the case of parameter space democracy (gray) and susy (black) boundary conditions.  The top line shows the width of the SM decay modes, $\Gamma_{\text{SM}}$.  The bottom line is $0.1 \Gamma_{SM}$. \\}
\label{HiggsDecay}
\end{figure}

It is possible that the scalars of the Higgs \textbf{5}-plet were produced at the $e^+$$e^-$ collider  LEP with $\sqrt{s} \sim 200$ GeV via the processes
\begin{equation} \label{LEP}
e^+ e^- \rightarrow \phi_5^+ \phi_5^-  \quad \text{and} \quad e^+ e^- \rightarrow a_5^0 s_5^0.
\end{equation}
LEP placed limits on the production cross sections for the neutralino and chargino \cite{Abbiendi:2004} and these bounds can be directly translated to the processes in (\ref{LEP}).  By doing so, approximate limits on the masses of the scalars $a_5, s_5$, and $\phi_5^{\pm}$ can be deduced.  The charged Higgs \textbf{5}-plet $\phi_5^{\pm}$ is ruled out for masses below $\sim 90$ GeV and the neutral scalar $s_5^0$ is ruled out for masses between $\sim 100 - 120$ GeV (depending on the mass of $a_5$).  Fig.~\ref{MassSplittings} summarizes the important constraints on the mass splittings $\Delta m_{\phi_5^{\pm} a^0}$ and $\Delta m_{s^0 a^0}$ for the light dark matter.     

The heavy dark matter candidate ($m_{a^0} \gtrsim 200$ GeV) could  not be produced at LEP.  In addition, its contributions to the electroweak parameters always fall within the $1 \sigma$ experimental bounds.  Thus, the main constraints on the mass splittings in this region of parameter space come from vacuum stability and perturbativity. Typically, $\Delta m_{\phi_5^{\pm} a^0}, \Delta m_{s^0 a^0} \lesssim 20$ GeV in the heavy dark matter regime.

%In particular, the cross section for production of the two lightest scalar five-plets is    
%%
%\begin{equation}
%\sigma = \frac{N_h}{6 \pi s^{5/2}} \Big(\frac{g}{2 \cos\theta_w}\Big)^4\frac{(\frac{1}{2}- 2 \sin\theta_w^2+4 \sin\theta_w^4)}{(1-m_z^2/s)^2} |\textbf{p$_f$}|^3,
%\end{equation}
%%
%where $|\textbf{p$_f$}|$ is the outgoing momentum in the center-of-mass frame.  For production of the charged five-plet, the cross section is 
%%
%\begin{eqnarray}
%\sigma &=& \frac{N_h s^{1/2} |\textbf{p$_f$}|}{12 \pi} \Big(1-\frac{4 m_{\phi^{\pm}}^2}{s}\Big) \nonumber \\
%&& \times \Big\{\Big[\frac{-e^2}{s}+B (4 \sin\theta_w^2-1)\Big]^2+B^2\Big\}
%\end{eqnarray}
%%  
%where $B= \frac{eg \cot 2 \theta_w}{4 \cos\theta_w} \frac{1}{s-m_z^2}$. 
One of the most promising discovery channels for the \textbf{5}-plet scalars at the Tevatron and LHC is the width of the SM Higgs.  The Higgs can decay into $a_5, s_5,$ or $\phi_5^{\pm}$, in addition to the SM modes. 
In Fig.~\ref{DarkMatter:Mass}, all points below the red dotted line satisfy $m_{h^0} > 2 m_{a^0}$; here, the SM Higgs can decay into the dark matter.  This decay channel is open for significant portions of both the parameter space democracy and susy boundary condition cases.  For small mass splittings, decays to $s_5$ and $\phi^{\pm}_5$ are also possible, though they are subdominant.

The contribution of the new invisible decays to the width of the Higgs is
\begin{eqnarray}
\Gamma_{\text{inv}} &=& \frac{N_h v^2}{32 \pi m_{h^0}} \Bigg[ \lambda_{\eff}^2 \sqrt{1-\frac{4 m_{a^0}^2}{m_{h^0}^2}}  + 2 \lambda_3^2 \sqrt{1-\frac{4 m_{\phi^{\pm}}^2}{m_{h^0}^2}} \nonumber \\
&& + (\lambda_{\eff} + 4 |\lambda_5|)^2 \sqrt{1-\frac{4 m_{s^0}^2}{m_{h^0}^2}} \Bigg]. 
\end{eqnarray}

Fig.~\ref{HiggsDecay} plots $\Gamma_{\text{inv}}$ as a function of the SM Higgs mass.  The (top) line is the width due to the SM decay modes \cite{Djouadi:2005gi}.  For points above the line, the invisible decays into the \textbf{5}-plet scalars are the dominant contribution.  However, even when  $\Gamma_{\text{inv}}\sim 0.1 \Gamma_{\text{SM}}$, it should be possible to detect the additional decay modes at the Tevatron or the LHC.  

The scalars may also be produced at the Large Hadron Collider via interactions like
\begin{eqnarray} \label{LHCproduction}
pp &\rightarrow& a_5^0 s_5^0 \rightarrow Z^* + \displaystyle{\not}{E_T}  \nonumber \\
pp &\rightarrow& \phi_5^+ \phi_5^-  \rightarrow W^* W^* + \displaystyle{\not}{E_T}   \nonumber \\
pp &\rightarrow& s_5^0 \phi_5^{\pm} \rightarrow Z^* W^* +\displaystyle{\not}{E_T} \nonumber \\
pp &\rightarrow& a_5^0 \phi_5^{\pm}  \rightarrow W^* + \displaystyle{\not}{E_T}. 
\end{eqnarray}
The vector bosons are always off-shell because the scalar mass splittings are less than 80 GeV (see Fig.~\ref{MassSplittings}) and, after using their leptonic branching fraction, it will be challenging to detect this signal in the presence of a large background.

\begin{figure}[b]
\includegraphics[width=3.3 in]{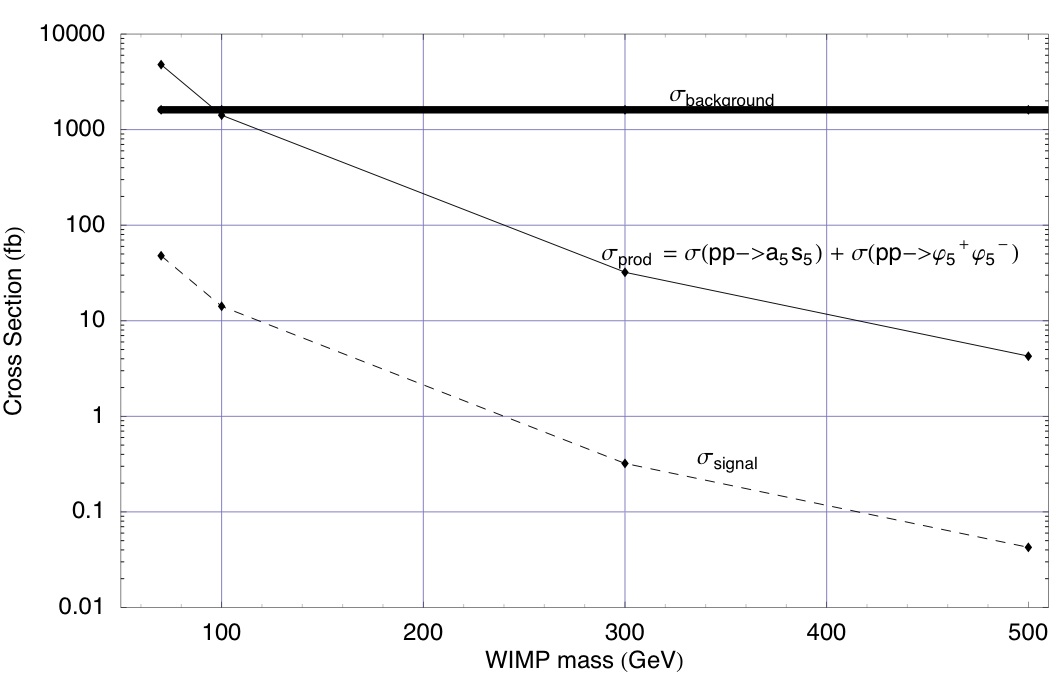}
\caption{LHC production cross section $\sigma_{\text{prod}}$ for $a_5^0 s_5^0$ and $\phi_5^+ \phi_5^-$, assuming $\Delta m_{\text{s$^0$a$^0$}}$ = 10 GeV, $\Delta m_{\text{$\phi^{\pm}$a$^0$}} = 15$ GeV, $m_{h^0} = 120$ GeV, and $\lambda_3 = 0.3$.  The dotted line is the signal cross section.  The thick black line is the cross section for the SM background (see text). }
\label{LHC}
\end{figure}

As an example, consider the first two processes in (\ref{LHCproduction}), which both result in opposite-sign leptons plus $\displaystyle{\not}{E_T}$ after the decay of the gauge bosons.  The production cross section $\sigma_{\text{prod}}$ for $s_5 a_5$ and $\phi_5^+ \phi_5^-$ at the LHC was calculated using MadGraph \cite{Maltoni:2003} for a sample point in parameter space and is plotted as a function of WIMP mass in Fig.~\ref{LHC}.    The cross section for the SM background (thick line) is
\begin{eqnarray}
\sigma_{\text{background}} &=& \sigma(pp \rightarrow WW) \text{Br}(W \rightarrow l \nu)^2 \nonumber\\
&&+\sigma(pp \rightarrow ZZ) \text{Br}(Z \rightarrow l^+ l^-) \text{Br}(Z \rightarrow \nu \nu). \nonumber \\
\end{eqnarray}
The signal cross section may be estimated as
\begin{equation}
\sigma_{\text{signal}} \sim \text{Br}(Z \rightarrow l^+ l^-) \sigma _{\text{prod}},
\end{equation}
 where the branching fraction is about 1\% (dashed line).  The ratio of signal to background is about 1:10 for the low mass dark matter.  At higher mass, it is about 1:1000.  This estimate indicates that it may be possible to see the signal for the low-mass DM region, if appropriate cuts are placed.  

 % The difficulty of detecting such scalars at the LHC has been discussed in \cite{Barbieri:2006}.  In \cite{Pierce:2007}, a modification of the inert doublet model is proposed so that results in the addition of 5 other heavy particles charged under $SU(3)_C$.  The presence of new colored particles increases the number of signals that one might expect to see at the LHC.

\section{Conclusions} \label{Discussion}

In this paper, a minimal extension of the Standard Model was presented that lead to gauge unification and a dark matter candidate.  An electroweak doublet $H_i$ in a {\bf 5} of a global discrete symmetry was introduced.  One of the new five-plet of particles is light and neutral and serves as a good dark matter candidate.  The addition of six scalars to the SM leads to gauge coupling unification and fixes the number of electroweak doublets.  

The six Higgs doublet model has distinct signatures for direct detection, indirect detection, and collider experiments.  Typically, the light mass range ($m_{a_5^0} \sim 80$ GeV) has the most promising signals, and will be tested for by GLAST and CDMS.  In addition, it can be produced by decays of the SM Higgs at the Tevatron or LHC.  The heavier candidates ($m_{a_5^0} \gtrsim 200$ GeV) are more difficult to see, but lie within the sensitivity of the HESS gamma ray detector and the next-generation direct detection experiment, SuperCDMS.  Direct production of these heavier candidates at colliders is challenging due to large Standard Model backgrounds from di-boson production, though further study is needed to determine whether appropriate cuts can reduce these backgrounds.

Throughout this discussion, it has been assumed that an exact discrete symmetry exists to keep the full five-plet of dark matter light under one fine-tuning.  Discrete symmetries can arise in string theoretic constructions (e.g., see \cite{DeWolfe:2004}), and the existence of these symmetries is critical for viability of this particular model.  
If this requisite symmetry is relatively common, then a single fine-tuning of the scalar mass is comparable to the tuning necessary in the ``minimal model'' described in \cite{Mahbubani:2006, Cirelli:2006}.  
In general, this class of minimal models is not as economical  in terms of fine-tuning as split susy, where the desired mass spectrum is obtained by having the R-symmetry breaking scale be small.  However, these minimal models give rise to interesting phenomenology that will be tested in upcoming experiments.

\begin{acknowledgments}
We would like to thank Roni Harnik and Ryuichiro Kitano for collaboration at the beginning of this work.  We would like to thank S. Dimopoulos, D. E. Kaplan,  and A. Pierce for useful conversations related to this work.    We would like to thank M. Tytgat for clarifications on the inert doublet model.  JGW and ML are supported under  the DOE under contract DE-AC03-76SF00515 and partially by the NSF under grant PHY-0244728.  ML is supported by an NDSEG fellowship.

\end{acknowledgments}

  \newpage

  \end{document}